\documentclass[twocolumn,aps,prd]{revtex4}
 \usepackage{graphicx,amssymb}
 \usepackage[utf8]{inputenc} % Required for inputting international characters
 
\usepackage[urlcolor=blue,colorlinks,breaklinks=true]{hyperref}

\PassOptionsToPackage{obeyspaces,spaces,hyphens}{url}
 
 \usepackage{amsmath}

\begin{document}
 	%My commands
 	\def\half{{1\over2}}
 	\def\shalf{\textstyle{{1\over2}}}
 	
 	\newcommand\lsim{\mathrel{\rlap{\lower4pt\hbox{\hskip1pt$\sim$}}
 			\raise1pt\hbox{$<$}}}
 	\newcommand\gsim{\mathrel{\rlap{\lower4pt\hbox{\hskip1pt$\sim$}}
 			\raise1pt\hbox{$>$}}}

\newcommand{\be}{\begin{equation}}
\newcommand{\ee}{\end{equation}}
\newcommand{\bq}{\begin{eqnarray}}
\newcommand{\eq}{\end{eqnarray}}
 	
%\title{Violation of Boltzmann's $H$ theorem in theories of gravity with nonminimally coupled matter fields}

\title{Reply to note on ``Boltzmann's H-theorem, entropy and the strength of gravity in theories with a nonminimal coupling between matter and geometry"}

\author{P.P. Avelino}
\email[Electronic address: ]{pedro.avelino@astro.up.pt}
\affiliation{Instituto de Astrof\'{\i}sica e Ci\^encias do Espa{\c c}o, Universidade do Porto, CAUP, Rua das Estrelas, PT4150-762 Porto, Portugal}
\affiliation{Departamento de F\'{\i}sica e Astronomia, Faculdade de Ci\^encias, Universidade do Porto, Rua do Campo Alegre 687, PT4169-007 Porto, Portugal}
\affiliation{School of Physics and Astronomy, University of Birmingham,Birmingham, B15 2TT, United Kingdom}

\date{\today}
\begin{abstract}
In this brief reply we respond to the note of Bertolami and Gomes (arXiv:2005.03968) on our recent paper (arXiv:2003.10154).
\end{abstract}

\maketitle

In \cite{avelino:2020fek} we have shown that Boltzmann's $H$-theorem does not necessarily hold in the context of theories of gravity with a nonminimal coupling (NMC) between the gravitational and matter fields. We have found sufficient conditions for the violation of Boltzmann's $H$-theorem, and derived an expression for the evolution of Boltzmann's $H$ in terms of the nonminimal coupling function ($\mathcal F (R)$, where $R$ is the Ricci scalar), valid in the case of a collisionless gas in a homogeneous and isotropic Friedmann-Lema\^itre-Robertson-Walker (FLRW) universe. We have highlighted the implications of this result for the evolution of the entropy, briefly discussing the role played by collisions between particles whenever they are relevant. We have also suggested a possible link between the high entropy of the Universe and the weakness of gravity in the context of these theories.\\

Our recent paper \cite{avelino:2020fek} agrees with, reinforces and extends our previous work on the subject \cite{Avelino:2018rsb,Avelino:2018qgt,Azevedo:2018nvi,Azevedo:2019krx}, and has never been intended as a note on  \cite{Bertolami:2020ldj} (despite disagreeing with its conclusions). In the most recent version of \cite{avelino:2020fek} we have made our best effort to give a more detailed account of aspects which might not have been sufficiently clear, providing alternative derivations of two of the main results of the paper. In the following we respond to the criticisms to our paper made in  \cite{Bertolami:2020fbf}. \\

\noindent {\bf Force on Point Particles (section III of \cite{avelino:2020fek})}\\

The authors of \cite{Bertolami:2020fbf} claim that section III of \cite{avelino:2020fek} copies without quoting, the discussion of \cite{Bertolami:2007gv}. This is not true, as can easily be checked by comparing references \cite{avelino:2020fek} and \cite{Bertolami:2007gv}. As acknowledged in \cite{avelino:2020fek}, the first part of section III concerns the determination of the 4-acceleration on point particles in the presence of a NMC to gravity and follows closely an analogous (standard) calculation done in \cite{Ayaita:2011ay} in the context of growing neutrino models where the neutrino mass is non-minimally coupled to a dark energy scalar field.\\

\noindent {\bf A note on the 4-acceleration of a fluid element (section IIIA of \cite{avelino:2020fek})}\\

The authors of \cite{Bertolami:2020fbf} claim that section IIIA of \cite{avelino:2020fek} is a repetition of the discussion of \cite{Bertolami:2008ab}. It is not, as can be easily verified by confronting references \cite{avelino:2020fek} and \cite{Bertolami:2008ab}. We acknowledge, however, that the original references \cite{Bertolami:2007gv,Bertolami:2008ab} should be quoted before Eq. (15), rather than a subsequent one \cite{Bertolami:2008zh}, of the same year and including the same authors, where that equation was also presented (we have rectified this in the most recent version of  \cite{avelino:2020fek}, correcting a typographical sign error found in both \cite{Azevedo:2019krx}, \cite{Bertolami:2008ab}, and \cite{Bertolami:2008zh}).\\

Section IIIA reminds the reader that, except in the case of dust, the 4-acceleration of the individual particles is in general not the same as the 4-acceleration of the fluid. This is essential for understanding the momentum-dependent forces --- damping or driving, depending, respectively, on whether the NMC function $\mathcal F(R)$ grows or decays with the cosmic time --- in the context of theories of gravity with a NMC between geometry and matter. \\

The authors of \cite{Bertolami:2020fbf} also claim that the on-shell Lagrangian of the matter fields $\mathcal L_{\rm m}=T$, where $T$ is the trace of the energy-monentum tensor, is unsuitable. This is again not true. In the case of a fluid made of many point particles the appropriate on-shell Lagrangian is indeed $\mathcal L_{\rm m}=T$ as demonstrated in \cite{Avelino:2018qgt,Avelino:2018rsb} (see also \cite{Ferreira:2020fma} for a detailed discussion of the appropriateness of the use of different Lagrangians to describe various components of the cosmic energy budget). In the most recent version of \cite{avelino:2020fek} we have expanded this subsection in order reinforce this point.\\

\noindent {\bf Boltzmann's $H$-Theorem (section IV of \cite{avelino:2020fek})}\\

The authors of \cite{Bertolami:2020fbf} argue that Boltzmann’s $H$-Theorem concerns the deviation from the incompressibility of the distribution function in phase space due to the existence of collisions and that the collisionless version of the Boltzmann equation implies that the distribution function is the equilibrium one concerning hence to adiabatic processes. Although this is true in the absence of momentum-dependent forces, it no longer holds if there is an extra momentum-dependent contribution to the variation of the linear momentum of the particles due to the NMC to gravity.\\

In the most recent version of \cite{avelino:2020fek} we have expanded this section in order to reinforce this point, and added a subsection providing an alternative derivation of how the evolution of Boltzmann's $H$ depends on the evolution of the NMC coupling function $\mathcal F (R)$ in a FLRW  universe in the absence of collisions (this is the main result of section IV of \cite{avelino:2020fek} and one of the most important of the paper). The addition of a collision term is discussed in section VA \cite{avelino:2020fek} (this discussion has been expanded and promoted to a subsection in the most recent version of the paper).\\

\noindent {\bf Entropy (section V of \cite{avelino:2020fek})}\\

The authors of \cite{Bertolami:2020fbf} claim that this section follows, without quoting, the remarks of  \cite{Bertolami:2020ldj} concerning Gibbs’ and Boltzmann’s $H$-theorem. This is not true, as can be easily checked by confronting references \cite{avelino:2020fek} and \cite{Bertolami:2020ldj}.\\

\noindent {\bf The Strength of Gravity (section VB of \cite{avelino:2020fek})}\\

The authors of \cite{Bertolami:2020fbf} remark that in \cite{avelino:2020fek} we assume that $\mathcal F_1(R) = R$, dropping the terms in the field equations that characterize the more general non-minimal curvature-matter coupling gravity theories (NMCCMGT) where the function $\mathcal F_1$ is a function of the Ricci scalar $R$. This is true. In the most recent version of the paper we added a note after Eq.~(1) clarifying that the generalization of the Lagrangian from $\mathcal L = R+ \mathcal F(R)\mathcal{L}_{\rm m}$ to  $\mathcal L = \mathcal F_1(R)+ \mathcal F_2(R)\mathcal{L}_{\rm m}$ is straightforward, and that the former is considered for simplicity (not affecting our main results).\\

The authors of \cite{Bertolami:2020fbf} also claim that the contra-intuitive situation in which the gravitational interaction could be stronger at early and late times is assumed, which would seem  to hint for a collapsing universe. This is not the case  considered in \cite{avelino:2020fek}. In the most recent version of \cite{avelino:2020fek} we have greatly simplified this section --- including it as a subsection of section V (section VB) --- hopefully clarifying the doubts raised in \cite{Bertolami:2020fbf}.\\

\noindent {\bf Other Issues}\\

The authors of \cite{Bertolami:2020fbf} claim that in \cite{avelino:2020fek} we consider a single particle approach in the NMC gravity and analyse a non-relativistic version of the collisionless Boltzmann equation to tackle relativistic considerations to surprisingly conclude that entropy variation arises from a collisionless Boltzmann equation. This is again not true. The collisionless case (relevant, for example in the case of neutrinos and photons after neutrino and photon decoupling, respectively) is treated in section IV of  \cite{avelino:2020fek}. To this end, the continuity equation, describing particle number conservation in six-dimensional phase space in the absence of collisions, has been used without making any assumptions regarding the relativistic or non-relativistic nature of the particles. The computation of the  momentum-dependent forces on the particles, covariantly defined in section III of  \cite{avelino:2020fek}, was crucial to solve the phase space continuity equation and to determine the  evolution of Boltzmann's $H$ in terms of the NMC coupling function $\mathcal F(R)$ in a FLRW universe. The impact of the addition of a collision term is briefly discussed in section VA of  \cite{avelino:2020fek}.\\

\noindent {\bf Final Remark}\\

We welcome and value criticisms to our work and will always be prepared to revise it in the light of valid scientific arguments or to provide a more detailed account of any aspects which might not have been sufficiently clear.

\bibliography{reply}
 	
 \end{document}